\newpage
\documentstyle[preprint,aps]{revtex}
\input epsf.tex
\def\DESepsf(#1 width #2){\epsfxsize=#2 \epsfbox{#1}}
\begin{document}
\preprint{\vbox{\hbox{}}}
\draft
\title{
$SU(3)$ Analysis of Annihilation Contributions and CP Violating Relations 
in $B\to P P$ 
Decays}
\author{Xiao-Gang He}
\address{
Department of Physics, National Taiwan University, Taipei, Taiwan 10764, R.O.C.
}
\date{October 1998}
\maketitle
\begin{abstract}
Several methods proposed to measure the angle $\gamma$ in the KM unitarity
triangle assumed that the tree contribution to $B^-\to \pi^-\bar K^0 $ is 
purely due to annihilation contributions
and is negligibly small. This assumption
has to be tested in order to have a correct interpretation of the experimental
data. In this paper we 
show that using SU(3) symmetry, the smallness of the tree contribution
can be tested in a dynamic model independent way. We also
derive several  relations between 
CP violating rate differences for $B\to P P$ decays without assuming the 
smallness of the annihilation contributions. These relations provide important 
tests for the Standard Model of CP violation.

\end{abstract}

\newpage
\section{Introduction}
Several rare two body decay modes of $B_{u,d}$ mesons have been observed 
at CLEO\cite{1}.
These data have 
provided 
interesting information about the Standard Model (SM)\cite{3,4,5,6}.
With increased luminosities for B-factories at CLEO, 
KEK and SLAC, more
useful information about rare $B_{u,d}$ decays will be obtained. 
The SM will be tested in detail.
At present the study of rare $B_s$ decays are limited by statistics. Only 
some weak upper limits on the branching ratios have been obtained\cite{2}. 
However, more data on $B_s$ decays will become available from 
LHC in the future.
These data
will help to further test the SM\cite{3,2a}.
Theoretical predictions are, however, limited by
our inability to reliably calculate many hadronic matrix elements related to
$B$ decays. 
This prevents a full test of the Standard Model.
In the lack of reliable calculations, attempts have
been made to extract useful information from symmetry considerations.
$SU(3)$ flavor symmetry is one of the symmetries which has attracted a lot of
attentions recently\cite{7,7a,7b}.
For example, it has been shown that using SU(3) symmetry it is possible to 
constrain\cite{8} and to 
determine\cite{5,9}
one of the fundamental parameters $\gamma$ in the SM
for CP violation by measuring several B meson decay modes. 

Some of the 
methods proposed to measure $\gamma$ depend on the assumption that the
tree amplitude to $B^-\to \pi^-  \bar K^0$
is negligibly small\cite{7b,9}. 
To correctly interpret the experimental data, the
smallness of the  tree contribution 
has to be confirmed experimentally.
It is often assumed that the tree amplitude for $B^-\to \pi^- \bar K^0$ 
receives annihilation contributions only. If this is true, one has to
make sure that these contributions are small. Of course one has to 
make sure that it is true that the decay amplitude is dominated by 
annihilation contributions.
There have 
been several discussions about constraining the annihilation
contributions using SU(3) analysis\cite{10}. In this paper we will 
use SU(3) symmetry to study further related problems,
but look at the problems in a different angle. We will 
first show how one can use SU(3) relations to test the smallness of 
annihilation contributions.  We then show that the statement that 
the tree amplitude receives annihilation contributions only
for $B^-\to \pi^- \bar K^0 $
is not  strictly a SU(3) result. 
We will show how to verify the smallness of the tree amplitude for 
$B^-\to \pi^- \bar K^0$ using 
several B decay modes.
Finally
we will use SU(3) symmetry to derive several useful relations regarding
CP violating rate differences without any assumption about the size of the 
annihilation contributions. These relations provide further tests for the
SM of CP violation and also the SU(3) symmetry.

\section{$SU(3)$ decay amplitudes for $B\to PP$}  

The quark level effective Hamiltonian up to one loop level in 
electroweak interaction 
for hadronic charmless $B$ decays, 
including the corrections to the matrix elements, can be written as 
\begin{eqnarray}
 H_{eff}^q = {4 G_{F} \over \sqrt{2}} [V_{ub}V^{*}_{uq} (c_1 O_1 +c_2 O_2) 
   - \sum_{i=3}^{12}(V_{ub}V^{*}_{uq}c_{i}^{uc} +V_{tb}V_{tq}^*
   c_i^{tc})O_{i}].
\end{eqnarray} 
The operators are defined in Ref.\cite{11}. The coeffecients
$c_{1,2}$ and $c_i^{jk}=c_i^j-c_i^k$, with $j$ indicates the internal quark, 
are the Wilson Coefficients (WC). These 
WC's have been evaluated by several groups\cite{11},
with $|c_{1,2}|>> |c_i^j|$.
In the above the factor $V_{cb}V_{cq}^*$ has 
been eliminated using the unitarity property of the KM matrix.

At the hadronic level, the decay amplitude can be generically written as

\begin{eqnarray}
A = <final\;state|H_{eff}^q|B> = V_{ub}V^*_{uq} T(q) + V_{tb}V^*_{tq}P(q)\;,
\end{eqnarray}
where $T(q)$ contains contributions from 
the $tree$ as well as $penguin$ due to charm and up 
quark loop corrections to the matrix elements, 
while $P(q)$ contains contributions purely from 
$penguin$ due to top and charm loops. 
The relative strength of the amplitudes
$T$ and $P$ is predominantly determined by their corresponding WC's in the 
effective Hamiltonian.
For $\Delta S = 0$ charmless decays, the dominant contributions are due to the
tree operators $O_{1,2}$ and the penguin operators are suppressed by smaller
WC's. Whereas for $\Delta S =-1$ decays, because the penguin contributions are 
enhanced by a factor of $V_{tb} V_{ts}^*/V_{ub}V_{us}^*\approx 55$ 
compared with the 
tree contributions, penguin effects dominate the 
decay amplitudes. In this case the electroweak penguins 
can also play a very important role\cite{13}, 
in particular when study CP violation in 
B decays\cite{14}. One should carefully keep track of the different
contributions.

The operators $O_{1,2}$, $O_{3-6, 11,12}$, and $O_{7-10}$ transform under SU(3)
symmetry as $\bar 3_a + \bar 3_b +6 + \overline {15}$,
$\bar 3$, and $\bar 3_a + \bar 3_b +6 + \overline {15}$, respectively. 
These properties enable us to 
write the decay amplitudes for $B\to PP$ in only a few SU(3) invariant 
amplitudes.

For the $T(q)$ amplitude, for example, we have\cite{7a} 
\begin{eqnarray}
T(q)&=& A_{\bar 3}^TB_i H(\bar 3)^i (M_l^k M_k^l) + C^T_{\bar 3}
B_i M^i_kM^k_jH(\bar 3)^j \nonumber\\
&+& A^T_{6}B_i H(6)^{ij}_k M^l_jM^k_l + C^T_{6}B_iM^i_jH(6
)^{jk}_lM^l_k\nonumber\\
&+&A^T_{\overline {15}}B_i H(\overline {15})^{ij}_k M^l_jM^k_l +
C^T_{\overline
{15}}B_iM^i_j
H(\overline {15} )^{jk}_lM^l_k\;,
\end{eqnarray}
where $B_i = (B_u,  B_d,  B_s) = (B^-, \bar B^0, \bar B^0_s)$ 
is a SU(3) triplet, $M_{i}^j$ is the SU(3) pseudoscalar 
octet, and the matrices $H(i)$ 
contain information about 
the transformation properties of the operators $O_{1-12}$.

For $q=d$, the non-zero entries of the matrices $H(i)$ are given by
\begin{eqnarray}
H(\bar 3)^2 &=& 1\;,\;\;
H(6)^{12}_1 = H(6)^{23}_3 = 1\;,\;\;H(6)^{21}_1 = H(6)^{32}_3 =
-1\;,\nonumber\\
H(\overline {15} )^{12}_1 &=& H(\overline {15} )^{21}_1 = 3\;,\; H(\overline
{15} )^{22}_2 =
-2\;,\;
H(\overline {15} )^{32}_3 = H(\overline {15} )^{23}_3 = -1\;.
\end{eqnarray}
And for $q = s$, the non-zero entries are
\begin{eqnarray}
H(\bar 3)^3 &=& 1\;,\;\;
H(6)^{13}_1 = H(6)^{32}_2 = 1\;,\;\;H(6)^{31}_1 = H(6)^{23}_2 =
-1\;,\nonumber\\
H(\overline {15} )^{13}_1 &=& H(\overline {15} ) ^{31}_1 = 3\;,\; H(\overline
{15} )^{33}_3 =
-2\;,\;
H(\overline {15} )^{32}_2 = H(\overline {15} )^{23}_2 = -1\;.
\end{eqnarray}

Due to the anti-symmetric property of $H(6)$ in exchanging the upper two indices,
$A_6$ and $C_6$ are not independent\cite{7a}. For individual decay 
amplitude, $A_6$ and $C_6$  always appear together in the form $C_6-A_6$. 
We will absorb $A_6$ in 
the definition of $C_6$.
In terms of the SU(3) invariant amplitudes, the decay amplitudes  for 
various B meson decays are given by

\begin{eqnarray}
\begin{array}{ll}
\Delta S = 0 &\Delta S = -1\\
T^{B_u}_{\pi^-\pi^0}(d) = {8\over \sqrt{2}}C^T_{\overline {15}},
&T^{B_u}_{\pi^-\bar K^0}(s)= C^T_{\bar 3}
 - C^T_{6} + 3A^T_{\overline {15}} -  C^T_{\overline {15}
 },\\
T^{B_u}_{\pi^- \eta_8}(d)={2\over \sqrt{6}}
(C^T_{\bar 3} - C^T_6 + 3 A^T_{\overline {15}} + 3C_{\overline {15}}),
&T^{B_u}_{\pi^0K^-}(s)= {1\over \sqrt{2}} (C^T_{\bar 3}
  - C^T_{6} + 3A^T_{\overline {15} } +7 C^T_{\overline {15}
  })\;,\\
T^{B_u}_{K^- K^0}(d)=
C^T_{\bar 3} - C^T_6 + 3 A^T_{\overline {15}} -C^T_{\overline {15}},
&T^{B_u}_{\eta_8K^-}(s)= {1\over\sqrt{6}}(-C^T_{\bar 3}
   + C^T_{6} - 3A^T_{\overline {15}} +9 C^T_{\overline {15}
   }),\\
T^{B_d}_{\pi^+\pi^-}(d) = 2A^T_{\bar 3} +C^T_{\bar 3}
 + C^T_{6} + A^T_{\overline {15} } + 3 C^T_{\overline {15}
},
&T^{B_d}_{\pi^+ K^-}(s) =  C^T_{\bar 3}
 + C^T_{6} - A^T_{\overline {15}} + 3 C^T_{\overline {15}
},\\
T^{B_d}_{\pi^0\pi^0}(d)= {1\over \sqrt{2}} (2A^T_{\bar 3} +C^T_{\bar 3}
 + C^T_{6} + A^T_{\overline {15} } -5 C^T_{\overline {15}
}),
&T^{B_d}_{\pi^0\bar K^0}(s)= -{1\over \sqrt{2}} (C^T_{\bar 3}
 + C^T_{6} - A^T_{\overline {15} } -5 C^T_{\overline {15} }),\\
T^{B_d}_{K^- K^+}(d)= 2(A^T_{\bar 3}  +  A^T_{\overline {15}}),
&T^{B_d}_{\eta_8 \bar K^0}(s)=  -{1\over \sqrt{6}} (C^T_{\bar 3}
 + C^T_{6} - A^T_{\overline {15} } -5 C^T_{\overline {15} }),\\
T^{B_d}_{\bar K^0 K^0}(d)= 2A_{\bar 3} + 
C^T_{\bar 3} - C^T_6 - 3 A^T_{\overline {15}} - C_{\overline {15}},
& 
T^{B_s}_{\pi^+\pi^-}(s) = 2(A^T_{\bar 3} 
+ A^T_{\overline {15}}),\\
T^{B_d}_{\pi^0 \eta_8}(d)= {1\over \sqrt{3}}
(-C^T_{\bar 3} + C^T_6 + 5 A^T_{\overline {15}} + C_{\overline {15}}),
&
T^{B_s}_{\pi^0\pi^0}(s) = \sqrt{2}(A^T_{\bar 3} 
+ A^T_{\overline {15}}),\\
T^{B_d}_{\eta_8 \eta_8}(d)={1\over \sqrt{2}}
(2A_{\bar 3} + {1\over 3} C^T_{\bar 3} - C^T_6 
-A^T_{\overline {15}} + C_{\overline {15}}),
&
T^{B_s}_{K^+K^-}(s)= 2A^T_{\bar 3} +C^T_{\bar 3}
+ C^T_{6} + A^T_{\overline {15} } + 3 C^T_{\overline {15}
},\\
T^{B_s}_{ K^+ \pi^-}(d) =   
C^T_{\bar 3} + C^T_6 -  A^T_{\overline {15}} +3 C_{\overline {15}},
&
T^{B_s}_{K^0\bar K^0}(s)= 2A^T_{\bar 3} +C^T_{\bar 3}
- C^T_{6} -3 A^T_{\overline {15} } - C^T_{\overline {15}
},\\
T^{B_s}_{ K^0 \pi^0}(d) =  
-{1\over \sqrt{2}}(C^T_{\bar 3} + C^T_6 -  A^T_{\overline {15}} 
-5 C_{\overline {15}}),
&
T^{B_s}_{\pi^0\eta_8}(s)= {2\over \sqrt{3}}
( C^T_{6}
+2 A^T_{\overline {15}} - 2C^T_{\overline {15}
}),\\
T^{B_s}_{K^0 \eta_8}(d)=  
-{1\over \sqrt{6}}(C^T_{\bar 3} + C^T_6 -  A^T_{\overline {15}} 
-5 C_{\overline {15}}),
&T^{B_s}_{\eta_8\eta_8}(s)= \sqrt{2}(A^T_{\bar 3} +{2\over 3} C^T_{\bar 3}
- A^T_{\overline {15} } - 2C^T_{\overline {15}
}).
\end{array}
\nonumber
\end{eqnarray}

The amplitudes for $P(q)$ in terms of SU(3) 
invariant amplitudes can be obtained in a similar way. We will indicate 
the corresponding
amplitudes 
by $A^P_i$ and $C^P_i$.

Many analysis have been carried out using SU(3) classification of quark level
diagrams\cite{7b}. In most cases such an analysis will obtain the same results as 
the use of SU(3) invariant amplitudes. However, in some cases the
classification according to quark level diagrams without care would loss 
some vital information.
An interesting example is the tree amplitude for
$B^- \to \pi^- \bar K^0 $. Using quark level diagram analysis, when the 
annihilation contributions are neglected, the tree operators do not
contribute to this decay. This implies, in the SU(3) invariant amplitude 
language, that

\begin{eqnarray}
C^T_{\bar 3} - C^T_6  - C^T_{\overline{ 15}} = 0.
\end{eqnarray}
This, however, is not generally true
as has been confirmed by model calculations\cite{15,15a}.
In the quark level 
diagram classification, there are only four independent amplitudes whereas the 
general SU(3) invariant classification, there are five independent 
amplitudes\cite{7a}. 
Some information related to 
different combinations of quark level diagrams and their phases have
been lost in the naive quark level diagram analysis.
Specifically, four quark operators containing $\bar d \Gamma_1 d \bar q 
\Gamma_2 b$ and $\bar s \Gamma_1 s \bar q \Gamma_2 b$
 types of terms, where $\Gamma_i$ indicate appropriate Dirac
matrices,
appear in SU(3)
invariant amplitudes do not appear in the naive tree quark diagram
analysis.
For this reason, we
will use the SU(3) invariant amplitude to carry out our analysis.

\section{Test the smallness of annihilation contributions}

The amplitudes $A_{\bar 3, \overline  {15}}$  correspond to annihilation 
contributions. Here we refer the amplitudes with one of the light quark 
in the effective Hamiltonian 
corresponds to the light quark inside the B mesons to be annihilation 
amplitudes. 
The amplitudes $A_{\bar 3, \overline {15}}$ are annihilation amplitudes
can be understood by noticing that the light quark index in the
$B$ mesons are contracted with the Hamiltonian\cite{17}. 
The A and E type of
contributions in the quark diagram classification are linear combinations of 
$A_{\bar 3}$ and $A_{\overline{15}}$. 
It has been argued that these contributions are small 
based on model calculations\cite{7b}. 
At present the annihilation contributions can not be reliablely calculated. 
In
view of this, it is important to be able to 
test the smallness of the annihilation 
contributions experimentally.

In this section we show that using 
SU(3) relations, the size of the annihilation contributions 
can be measured independent dynamic models for the matrix 
elements and therefore the smallness of these amplitudes can be 
tested. Two types of tests
can be carried out. One of them is to test the smallness of the 
annihilation contributions of the SU(3) invariant amplitudes,
and another 
is to test the smallness 
of tree contribution to $B^-\to \pi^- \bar K^0$.

The best way to test the smallness of the annihilation contributions
is to use processes involving only $A_{\bar 3, \overline {15}}$. 
From discussions
of the previous section, we find that there are only three such processes.
They are: 
a) $\bar B^0 \to K^+ K^-$; b) $B_s\to \pi^-\pi^+$;
And c) $B_s\to \pi^0\pi^0$.
Their decay amplitudes are given by

\begin{eqnarray}
&&A(B_d\to K^+ K^-) = 
2V_{ub}V_{ud}^*(A^T_{\bar 3} + A^T_{\overline {15}})
+ 2V_{tb}V_{td}^*(A^P_{\bar 3} + A^P_{\overline {15}}),\nonumber\\
&&A(B_s\to \pi^- \pi^+)=
2V_{ub}V_{us}^*(A^T_{\bar 3} + A^T_{\overline {15}})
+ 2V_{tb}V_{ts}^*(A^P_{\bar 3} + A^P_{\overline {15}}),\nonumber\\
&&A(B_s\to \pi^0\pi^0) ={1\over \sqrt{2}} A(B_s\to \pi^+\pi^-).
\end{eqnarray}

It is clear that these decays receive annihilation contributions only.
However, there is a crucial difference between
a), and b) and c).
The decay amplitude for a) is dominated by the tree contribution
and the amplitudes for 
b) and c), being $\Delta S = -1$ processes, are dominated by 
penguin contributions. If annihilation contributions are small, these 
processes will all have small branching ratios. 
At present, 
these three modes have not been observed. The best constraint is from
$\bar B^0 \to K^+K^-$ with an upper bound on the branching ratio
$0.24\times 10^{-5}$ at the 90\% confidence level from CLEO\cite{1}.
However this still allow substantial annihilation contributions. The 
annihilation contributions to the tree amplitude to 
$B^-\to \pi^- \bar K^0 $ can reach 10\% of the total ampltidue. We have to
wait more data to verify the smallness of the annihilation contributions. 
Conclusions drawn with such assumption should be viewed with caution.

One should be aware that even the annihilation contributions are small, it 
does not mean that the tree amplitude for $B^-\to \pi^- \bar K^0$ is small. 
One has also to verify that the tree amplitude receives annihilation 
contributions only. Let us now study how this can be verified. 
From the SU(3) decay amplitudes listed 
in the previous section, we see that the tree contribution
to this process is given by

\begin{eqnarray}
T^{B_u}_{\pi^- \bar K^0 } (s) = C_{\bar 3}^T - C_6^T + 3A^T_{\overline {15}} 
- C^T_{\overline {15}}.
\end{eqnarray}
This is not a pure annihilation process as for 
$\bar B^0 \to K^+ K^-$, $B_s\to \pi^-\pi^+$, and
$B_s\to \pi^0\pi^0$. 
The tree amplitude to $B^-\to \pi^- \bar K^0 $ is pure annihilation 
contribution only in the 
factorization calculation 
where $C_{\bar 3}^T - C_6^T - C^T_{\overline {15}} = 0$.
In order this to be true, not only the magnitude of 
the invariant amplitudes should be arranged, but also the phases
of these amplitudes must be arranged to have the cancellation. 
However, from our experience with K and D systems, we know that 
different SU(3) (or isospin) amplitudes develop different phases.
It is 
quite possible that the same situation happens in B system\cite{16,16a}. 
To have a better understanding of the situation,
let us perform a calculation of the tree
decay amplitude for 
$T(B^-\to \bar K^0 \pi^-)$ in the factorization approximation 
neglecting the annihilation contributions, 
but with insertions of possible final state interaction
phases for different amplitudes.
We have\cite{4,14}

\begin{eqnarray}
&&T(B^-\to \pi^- \bar K^0 )=
V_{ub}V^*_{us}
(e^{i\delta_1}T_1 -e^{i\delta_3}T_3),\nonumber\\
&&T_1=T_3={1\over 3}
i{G_F\over  \sqrt{2}}
[(c_1 +{c_2\over N})
f_\pi F_0^{BK}(m^2_\pi) (m_B^2-m_K^2) + ({c_1\over N}+c_2)f_K F^{B\pi}_0
(m_K^2)(m_B^2-m_\pi^2)],
\end{eqnarray}
where 
 $N$ is the number of colors. 
We have used the following definitions  for the
decay constants and form factors
\begin{eqnarray}
&&<P|\bar q \gamma_\mu(1-\gamma_5) u|0> = if_P P_\mu\;,
\nonumber\\
&&<P(k)|\bar q \gamma_\mu b|\bar B^0(p)> = (k+p)_\mu
F^{BP}_1+(m_P^2-m_B^2){q_\mu\over q^2}(F^{BP}_1(q^2)
-F^{BP}_0(q^2)),
\end{eqnarray}
where $q = p-k$.
The first term $e^{i\delta_1}T_1$ 
in the amplitude $T(B^-\to \pi^-  \bar K^0)$ is 
equal to $C^T_{\bar 3} - C_6^T$ which is an $I = 1/2$ amplitude
while the second term $e^{i\delta_3} T_3$ is
equal to $C_{\overline {15}}^T$ which is an $I = 3/2$ amplitude. 
We see that the cancellation happens only when 
$\delta_1 = \delta_3$ which is an additional assumption about the
dynamics beyond SU(3) symmetry. 
It has been shown that present data does not exclude large final 
phase difference 
$\delta_1 - \delta_3$\cite{4,16a}.
The smallness of $C_{\bar 3} - C_6 - C_{\overline {15}}$ has 
to be tested experimentally. 

To have a model independent test
of this cancellation, that is, $C_{\bar 3} - C_6 -C_{\overline {15}} = 0$, 
one needs to find  processes which depend 
on the same combination of the SU(3) invariant amplitudes as the 
tree amplitude for $B^-\to \pi^- \bar K^0$. To this end we carry out 
an analysis similar to Refs. \cite{15a} 
for $B^-\to K^- K^0$
using the parametrization of the 
SU(3) decya amplitudes in the previous section.  We have

\begin{eqnarray}
A(B^-\to K^- K^0) = V_{ub}V_{ud}^* T^{B_u}_{K^- K^0}(d)
+ V_{tb}V_{td}^* P^{B_u}_{K^- K^0}(d).
\end{eqnarray}
As have been mentioned earlier that the relative strength of the 
T and P amplitudes is predominantly determined by their WC's,
to a good approximation 
$A(B^-\to K^- K^0) \approx V_{ub}V_{ud}^* T^{B_u}_{K^- K^0}(d)$.
In the SU(3) limit 

\begin{eqnarray}
T^{B_u}_{\pi^- \bar K^0}(s) = T^{B_u}_{K^- K^0}(d)
=C_{\bar 3}-C_6 +3 A_{\overline {15}}-C_{\overline {15}}.
\end{eqnarray}

Once the branching ratio for $B^- \to K^- K^0$ is measured, we have
information about the size of $|T^{B_u}_{\bar K^0 \pi^-}|$.
If experimentally, the branching ratio 
$B^-\to K^- K^0$ indeed turns out to be small,
this would confirm the smallness of $C_{\bar 3}-C_6-C_{\overline {15}}$ 
if annihilation contributions are also found to be small from the 
branching ratio measurements for $\bar B^0\to K^+ K^-$, $B_s\to \pi^+\pi^-$
and $B_s\to \pi^0\pi^0$.
In this case conclusions drawn with
the assumption, $T^{B_u}_{\pi^- \bar K^0 }(s) = 0$ would be good ones. 
Otherwise the results obtained with this assumption 
can not be trusted.
Unfortunately, at present experimental upper bound,
with $Br(B^-\to K^- K^0) < 0.93\times 10^{-5}$ at 90\% confidence level 
from CLEO\cite{1}, still allow large tree 
contributions to $B^-\to \pi^- \bar K^0$. 

We stress that the smallness for annihilation contributions and 
the smallness of the tree amplitude for
$B^-\to \pi^- \bar K^0$ are two independent assumptions and should be
tested separately as discussed in the above.
These tests have important implications for the
determination of the angle $\gamma$ in the KM unitarity triangle because
some of the methods proposed 
require that the tree amplitude is small such that 
$A(B^-\to \pi^- \bar K^0) = \bar A(B^+\to \pi^+ K^0)$. At present this is 
not well tested. 
We have to wait experiments in the future to tell us more. 

\section{CP asymmetry relation between $B$ decays}

From the previous discussions, we see that predictions with certain dynamic 
assumptions about the amplitudes suffer from possible uncertainties and
need to be tested. It is desirable that tests for the SM 
can be performed in a dynamic model independent way. 
In this section we will derive several such relations which can be used to 
test the Standard Model.
These relations are related to CP violating rate difference defined as

\begin{eqnarray}
\Delta(B\to PP) = \Gamma(B\to P P) - \Gamma(\bar B \to \bar P \bar P).
\end{eqnarray}

SU(3) symmetry relates $\Delta S = 0$ and 
$\Delta S =-1$ decays. One particularly
interesting class of relations are the ones with $T(d) = T(s)=T $ and
$P(d) = P(s) = P$. For this class of decays, we have\cite{17,19}

\begin{eqnarray}
A(d) = V_{ub}V_{ud}^* T + V_{tb}V_{td}^* P,\nonumber\\
A(s) = V_{ub}V_{us}^* T + V_{tb}V_{ts}^* P.
\end{eqnarray}
Due to different KM matrix elements involved in $A(d)$ and $A(s)$,
although the amplitudes have some similarities, the branching ratios are 
not simply related. However, when considering rate difference, $\Delta(
B\to PP)$, the situation is dramatically different. Because a simple property
of the KM matrix element\cite{20}, $Im (V_{ub}V_{ud}^*V_{tb}^*V_{td})
=-Im(V_{ub}V_{us}^*V_{tb}^*V_{ts})$, we find that in the SU(3) limit,

\begin{eqnarray}
\Delta(d) =- \Delta(s),
\end{eqnarray}
where $\Delta(i) = (|A(i)|^2-|\bar A(i)|^2) \lambda_{ab}/(8\pi m_B)$ is the
CP violating rate difference defined earlier and 
$\lambda_{ab} = \sqrt{1-2(m_a^2+m_b^2)/m_B^2 + (m_a^2-m_b^2)^2/m_B^4}$ with
$m_{a,b}$ being the masses of the two particles in the final state.

In the SU(3) limit we find the following equalities:

\begin{eqnarray}
1)&\;\;& 
\Delta(B^- \to K^- K^0) = - \Delta (B^- \to \pi^- \bar K^0)\;,\nonumber\\
2)&& \Delta(\bar B^0 \to \pi^- \pi^+) = - \Delta (B_s \to K^- K^+)\;,\nonumber\\
3)&& \Delta(\bar B^0 \to K^- K^+) = - \Delta (B_s \to \pi^- \pi^+)
\nonumber\\
&&=-2\Delta(B_s\to \pi^0\pi^0)\;,\nonumber\\
4)&& \Delta(\bar B^0 \to \bar K^0 K^0) = - \Delta (B_s \to K^0 \bar K^0)\;,\nonumber\\
5)&& \Delta(\bar B^0\to \pi^+ K^-) = - \Delta (B_s\to K^+\pi^-),\nonumber\\
6)&& \Delta(\bar B^0 \to \pi^0 \bar K^0) = - \Delta ( B_s \to K^0 \pi^0)
\nonumber\\
&&
=3\Delta(\bar B^0\to \eta_8\bar K^0) 
=-3\Delta(B_s\to K^0\eta_8).
\end{eqnarray}

Note that in the SU(3) limit, beside the above relations there are several 
other relations for the branching ratios, that is, some of the 
decay amplitudes are
actually equal in the SU(3) limit. We have

\begin{eqnarray}
&&\Gamma(B_s\to \pi^+\pi^-) = 2 \Gamma(B_s\to \pi^0\pi^0),\nonumber\\
&&\Gamma(\bar B^0\to \pi^0 \bar K^0) = 3 \Gamma(\bar B^0 \to \eta_8 \bar K^0),
\nonumber\\
&&\Gamma( B_s \to K^0 \pi^0) = 3\Gamma( B_s \to K^0 \eta_8 ).
\end{eqnarray} 

The last two equalities for the decay rate involve
$\eta_8$ which mixes with $\eta_1$.
It will be difficult to carry out these tests. The
branching ratio for the 
first one may be small due to pure annihilation contributions, although it 
has to be tested independently. This test will also be difficult to carry out. 

If it turns out that the annihilation contributions are all small as can be
tested in 
$B^-\to K^- K^0$, $B_s \to \pi^+\pi^-$ and $B_s\to \pi^0\pi^0$, 
there are additional relations for rate differences. We find

\begin{eqnarray}
&&1) \approx 4),\nonumber\\
&&2) \approx 5), \nonumber\\
&&6) \approx \Delta(\bar B^0 \to \pi^0\pi^0)
\end{eqnarray}
In the limit that annihilation contributions are small, it is difficult to 
perform tests related to 
1), 3) and 4) 
because the decay rates involved are all small.
The equalities of 
2) and 5) provide the best chances to test the SM.

The above non-trivial equalities do not
depend on the numerical values of the final state rescattering phases.
Of course these relations are true only for the SM 
with three generations. 
Therefore they provide  tests for the three generation
model.

The relations obtained above
hold in the SU(3) limit. Let us now study how
these relations are modified when SU(3) breaking effects are included.
Since no reliable calculational tool exists, 
in the following we will use factorization 
approximation neglecting the annihilation contributions 
to estimate the SU(3) breaking effects for
2) for illustration.
We have\cite{17}

\begin{eqnarray}
T^{B_d}_{\pi^-\pi^+}(d) &=& i{G_F\over \sqrt{2}}f_{\pi}F^{B\pi}_0(m_\pi^2)
(m_B^2-m_\pi^2)[{1\over N} c_1 +c_2
+{1\over N} c_3^{uc} +c_4^{uc} + {1\over N} c_9^{uc}
+c_{10}^{uc}\nonumber\\
&+& {2m_\pi^2 \over (m_b-m_u)(m_u+m_d)}
({1\over N} c_5^{uc} +c_6^{uc} +{1\over N} c_7^{uc} + c_8^{uc})]\;,\nonumber\\
T^{B_s}_{K^+ K^-}(s) &=& i{G_F\over \sqrt{2}}f_{K}F^{BK}_0(m_K^2)
(m_B^2-m_\pi^2)
[{1\over N} c_1 +c_2 +{1\over N} c_3^{uc} +c_4^{uc} +{1\over N} c_9^{uc}
+c_{10}^{cu}\nonumber\\
& +& {2 m_K^2 \over (m_b-m_u)(m_u+m_s)}
({1\over N} c_5^{uc} +c_6^{uc} +{1\over N} c_7^{uc} + c_8^{uc})]\;.
\end{eqnarray}
 The amplitudes
$P(d,s)$ are obtained by setting $c_{1,2} = 0$ and replacing 
$c_i^{uc}$ by $c_i^{tc}$.  

Using the fact $m_\pi^2/(m_u+m_d) \approx 
m_K^2/(m_u+m_s)$, we obtain
\begin{eqnarray}
\Delta(\bar B^0 \to \pi^+\pi^-)\approx 
-{(f_\pi F^{B\pi}_0(m_\pi^2))^2\over
(f_{K} F^{B_sK}_0(m_K^2))^2}{\lambda_{\pi\pi}\over \lambda_{K K}}
\Delta( B_s\to K^+K^-),
\end{eqnarray}
In the above the final state interaction phases 
for different amplitudes have been assumed to be zero. We point out that 
as long 
as these 
phases satisfy SU(3) symmetry relations, the above equation does not change.
 
Similarly we also have

\begin{eqnarray}
\Delta(\bar B^0 \to \pi^+\pi^-) &\approx&
-{(f_\pi F^{B\pi}_0(m_\pi^2))^2\over
(f_{K} F^{B\pi}_0(m_\pi^2))^2}{\lambda_{\pi \pi}\over \lambda_{\pi K}}
\Delta( \bar B^0\to \pi^+ K^-)
\nonumber\\
&\approx&{(f_\pi F^{B\pi}_0(m_\pi^2))^2\over
(f_{\pi} F^{B_sK}_0(m_\pi^2))^2}{\lambda_{\pi \pi}\over \lambda_{\pi K}}
\Delta( \bar B_s\to K^+\pi^-).
\end{eqnarray}

The form factors are 
usually assumed to have pole form  dependence on $q^2$. For the above cases
the form factors are approximately equal to their values at $q^2=0$ because the
B meson mass is much larger than $\pi$ and $K$ meson masses. For the
same reason, $\lambda_{\pi\pi} /\lambda_{\pi K} \approx  1$.
Independent of the specific value for the ratio $ r = 
F^{B\pi}_0(0)/F^{B_s K}_0(0)$, we obtain the following relations:

\begin{eqnarray}
&&\Delta(\bar B^0\to \pi^+\pi^-) \approx - {f_\pi^2\over f_K^2}
\Delta(\bar B^0\to \pi^+ K^-),\nonumber\\
&&\Delta( B_s\to K^+ K^-) \approx - {f_K^2\over f_\pi^2}
\Delta( B_s\to \pi^- K^+).
\end{eqnarray}
The first equality in the above has already been obtained before\cite{17}.
The ratio $r$ is expected to be about one. If this is indeed the case, one
would obtain $\Delta(\bar B^0 \to \pi^+\pi^-) \approx 
\Delta(B_s \to K^+ \pi^-)$.

It has been shown that the normalized asymmetry, that is, the rate difference 
divided by the averaged particle and anti-particle branching for 
$\bar B^0 \to \pi^+ K^-$, can be as large as 20\%\cite{4,16a}. Such a large 
value can be measured in the future at B factories. The Standard Model can be 
tested using the relations discussed in this section.

\section{CONCLUSIONS and discussions}

Several methods proposed to measure the fundamental parameter $\gamma$ in the
KM unitarity triangle depend on the assumption that,
$A(B^-\to \pi^- \bar K^0)= 
\bar A(B^+\to K^0 \pi^+)$. In order this assumption to hold 
it is not sufficient to only require the annihilation contributions to be
small. One has also to show that 
the tree amplitude only 
receives annihilation contributions. 
In this paper we have shown that these two conditions can be separately 
tested at B factories in the near future. 
Of course 
one should also keep an open mind the possibility that the annihilation
contribution $A_{\overline{15}}$ is
not small, but the total tree contribution 
$C^T_{\bar 3} - C^T_6 +3 A^T_{\overline{15}} + C^T_{\overline{15}}$ is small.
This can also be tested by measuring 
$B^-\to K^- K^0$ branching ratio because the dominant contribution to 
the amplitude is proportional to the tree amplitude for 
$B^-\to \pi^- \bar K^0$.

We have also derived several
useful relations using SU(3) symmetry without any additional dynamic 
model assumptions about the amplitudes. These relations will provide further
tests for the Standard Model of CP violation. 
The SU(3) symmetry is expected to be broken in reality. Therefore the
validity about some of the methods for measuring 
$\gamma$ and the relations derived in this paper remain to be a problem to
be studied. 

Let us conclude with a discussion about the validity of SU(3) relations for
B meson decays.
We have used factorization approximation to provide some idea
about how the SU(3) breaking effects affect the results. We stress that these
results are only indicative. 
One should not exclude the possibility that the experimental results obtained
will be actually more closer to the SU(3) limit results. 
Even though we know that
SU(3) symmetry is broken in reality, the breaking pattern may be much more
subtle than a simple decay constant rescaling as indicated from our 
factorization calculations in previous sections.
To see why this might happen let us consider
$B^-\to D^0 \pi^-$ and $B^-\to D^0 K^-$ decays. 

We find that in the SU(3) limit
the ratio $R = Br(B^-\to D^0 K^-) /Br(B^-\to D^0 \pi^-)$ is 
equal to $|V_{us}/V_{ud}|^2 
(\lambda_{DK}/\lambda_{D\pi})$. The value $R = 0.049$ obtained in the
SU(3) limit
is more closer to the experimental central 
value of $0.055\pm 0.015\pm0.005$ from CLEO\cite{21} than the factorization 
estimate with SU(3) breaking 
$R \approx (f_K^2/f_\pi^2) |V_{us}/V_{ud}|^2(\lambda_{DK}/
\lambda_{D\pi}) \approx 0.07$. 
Of course the experimental result is consistent with both predictions at the
present. The point of this example is that 
one should be careful about factorization estimate of SU(3) breaking
effects. 
SU(3) relations may turn out to be better than expected. 
We have to wait more experimental data to provide us with more
information. 

The above discussion also applies to
the relation between the tree amplitude $A^T$  for $B^-\to \pi^- \pi^0$ and the 
$I = 3/2$ tree amplitude $A^T_{3/2}$ for $B^- \to \pi^0 K^-$ and 
$B^-\to \pi^- \bar K^0$ decays. The experimental 
value may turn out to be closer to the SU(3) limit result than the 
factorization estimated relation\cite{5,9}
$A^T_{3/2} = (f_K^2/f_\pi^2) |V_{us}/V_{ud}|^2
A^T$. This also have important implications for the 
determination of $\gamma$. Any method to determine $\gamma$ using this relation should be analyzed with care.

This work was supported in part by 
ROC National Science Council and by Australian Research Council. 
I thank Hai-Yang Cheng for discussions and Alex Kagan for discussions on 
Ref. [18].

\end{document}